\documentclass[12pt]{article}
\usepackage[english]{babel}
\usepackage{multind}
\usepackage{lineno} 
\usepackage{graphicx,multicol}
\usepackage{epic,eepic,epsfig}
\usepackage{caption}

\usepackage{amsmath,amsfonts,amsthm,amssymb}
\usepackage{pstricks,pstricks-add,pst-math,pst-xkey}

\usepackage{amsmath,amsfonts,amsthm,amssymb}
\usepackage{pstricks,pstricks-add,pst-math,pst-xkey}

\usepackage{algorithm, algorithmic}

\usepackage{graphicx,pstricks,pst-node,amssymb}

\usepackage{tikz}
\pgfdeclarelayer{edgelayer}
\pgfdeclarelayer{nodelayer}
\pgfsetlayers{edgelayer,nodelayer,main}

\tikzstyle{none}=[inner sep=0pt]
\definecolor{hexcolor0xf81e1c}{rgb}{0.973,0.118,0.110}
\definecolor{hexcolor0x3c00ff}{rgb}{0.235,0.000,1.000}

\tikzstyle{whitevertex}=[circle,fill=white,draw=black, scale = 0.5]
\tikzstyle{redvertex}=[circle,fill=hexcolor0xf81e1c,draw=black, scale = 0.5]
\tikzstyle{bluevertex}=[circle,fill=hexcolor0x3c00ff,draw=black, scale = 0.5]
\tikzstyle{greenvertex}=[circle,fill=green,draw=black, scale=0.5]
\tikzstyle{purplevertex}=[circle,fill=magenta,draw=black, scale=0.5]
\tikzstyle{grayvertex}=[circle,fill=white,draw=gray, scale=0.5]
\tikzstyle{blackvertex}=[circle,fill=black,draw=black, scale=0.5]

\tikzstyle{textbox}=[rectangle,fill=none,draw=none]
\tikzstyle{box}=[rectangle,fill=none,draw=black]

\tikzstyle{arc}=[black, ->]
\tikzstyle{grayarc}=[gray, ->]
\tikzstyle{bluearc}=[blue, ->]
\tikzstyle{grayedge}=[draw=gray]
\tikzstyle{blueedge}=[draw=blue]
\tikzstyle{rededge}=[draw=red]
\tikzstyle{edge}=[draw=black]

\tikzstyle{vertex}=[circle, ,fill=white,draw=black, scale=0.5]
\tikzstyle{10circle}=[circle, scale=10.0,draw=black]
\tikzstyle{10oval}=[ellipse, scale=10.0,draw=black]

\usepackage{amsmath,amsfonts,amsthm,amssymb}
\usepackage{pstricks,pstricks-add,pst-math,pst-xkey}

\usepackage{algorithm, algorithmic}

\usepackage{graphicx,pstricks,pst-node,amssymb}

\usepackage{algorithm, algorithmic}

\begin{document}
\date{}

\newtheorem{tm}{\hspace{5mm}Theorem}[section]
\newtheorem{prp}[tm]{\hspace{5mm}Proposition}
\newtheorem{dfn}[tm]{\hspace{5mm}Definition}
\newtheorem{lemma}[tm]{\hspace{5mm}Lemma}
\newtheorem{cor}[tm]{\hspace{5mm}Corollary}
\newtheorem{conj}[tm]{\hspace{5mm}Conjecture}
\newtheorem{prob}[tm]{\hspace{5mm}Problem}
\newtheorem{quest}[tm]{\hspace{5mm}Question}
\newtheorem{alg}[tm]{\hspace{5mm}Algorithm}
\newtheorem{sub}[tm]{\hspace{5mm}Algorithm}
\newcommand{\induce}[2]{\mbox{$ #1 \langle #2 \rangle$}}
\newcommand{\2}{\vspace{2mm}}
\newcommand{\dom}{\mbox{$\rightarrow$}}
\newcommand{\ndom}{\mbox{$\not\rightarrow$}}
\newcommand{\compdom}{\mbox{$\Rightarrow$}}
\newcommand{\cdom}{\compdom}
\newcommand{\sdom}{\mbox{$\Rightarrow$}}
\newcommand{\la}{\langle}
\newcommand{\ra}{\rangle}
\newcommand{\pf}{{\bf Proof: }}
\newtheorem{claim}{Claim}
\newcommand{\beq}{\begin{equation}}
\newcommand{\eeq}{\end{equation}}
\newcommand{\<}[1]{\left\langle{#1}\right\rangle}

\newcommand{\Z}{\mathbb{$Z$}}
\newcommand{\Q}{\mathbb{$Q$}}
\newcommand{\R}{\mathbb{$R$}}

%\bibliographystyle{plain}
%\pagewiselinenumbering
%\setpagewiselinenumbers
%\modulolinenumbers[1]
%\linenumbers

\title{End-vertices of LBFS of (AT-free) bigraphs}

\author{Jan Gorzny and Jing Huang\thanks{
         Department of Mathematics and Statistics, 
         University of Victoria, Victoria, B.C., Canada V8W 3R4; 
         huangj@uvic.ca}}

\maketitle

\begin{abstract}
Lexicographic Breadth First Search (LBFS) is one of fundamental graph search 
algorithms that has numerous applications, including recognition of graph classes, 
computation of graph parameters, and detection of certain graph structures. 
The well-known result of Rose, Tarjan and Lueker on the end-vertices of LBFS of 
chordal graphs has tempted researchers to study the end-vertices of LBFS of 
various classes of graphs, including chordal graphs, split graphs, interval graphs, 
and asteroidal triple-free (AT-free) graphs. 
In this paper we study the end-vertices of LBFS of bipartite graphs.
We show that deciding whether a vertex of a bipartite graph is the end-vertex of 
an LBFS is an NP-complete problem. 
In contrast we characterize the end-vertices of LBFS of AT-free bipartite graphs. 
Our characterization implies that the problem of deciding whether a vertex of 
an AT-free bipartite graph is the end-vertex of an LBFS is solvable 
in polynomial time. 
\end{abstract}


\section{Introduction}

In 1976, Rose, Tarjan and Lueker \cite{rtl} introduced a variant of the Breadth
First Search (BFS) called the {\em Lexicographic Breadth First Search} (LBFS). 
LBFS modifies the selection rule of BFS to selecting at each step of the search
a vertex whose neighbours among the visited vertices form a lexicographically 
the least recent set. This seemingly little modification has a surprising impact 
on the resulting vertex ordering of the input graph. 
As shown in \cite{rtl}, when the input graph is chordal, 
the vertex ordering produced by an LBFS (called an {\em LBFS ordering}) is 
a {\em perfect elimination ordering}. 
Since perfect elimination orderings exist only for chordal graphs, LBFS 
correctly recognizes chordal graphs and finds perfect elimination orderings 
whenever possible. With perfect elimination orderings of chordal graphs, 
the {\em basic optimization problems} (the maximum clique, 
the minimum colouring, the maximum independent set, and 
the minimum clique covering) can all be solved efficiently, cf. \cite{golumbic}.

The key for proving the ordering of a chordal graph produced by an LBFS is 
a perfect elimination ordering is to show that the last visited vertex 
(called the {\em end-vertex}) of an LBFS is a {\em simplicial vertex}.
The characteristic 4-point property of LBFS orderings (cf. \cite{bdn,ck}) and 
the fact that the input graphs contains no induced cycles of length four or more
are central to the proof. The beautiful 4-point property of LBFS orderings and 
the elegant description of the end-vertices of LBFS of chordal graphs have tempted 
researchers to explore LBFS orderings and end-vertices of other graphs, 
cf.  \cite{bbbs,bb,chk,chm,cor,ckl,dragan,cos99,cos09,dn}.

Corneil, Olariu and Stewart \cite{cos99,cos09} studied the end-vertices of 
LBFS of {\em Asteroidal Triple-free} (AT-free) graphs and the end-vertices
of interval graphs. They showed that the end-vertex of an LBFS of an AT-free graph 
is {\em admissible}, cf. \cite{cos99}. 
Since interval graphs are chordal and AT-free, the end-vertex of an LBFS of 
an interval graph is both simplicial and admissible. 
It turns out that the converse is also true. Thus the end-vertices of LBFS of
interval graphs are precisely those which are both simplicial and admissible, 
cf. \cite{cos09}.  

Theoretical and algorithmic results on end-vertices of LBFS of (general) graphs 
are obtained in \cite{bb,ckl}. It is shown in \cite{bb} that the end-vertex
of an LBFS of an arbitrary graph must be in a {\em moplex} (which is a clique
module whose neighbourhood is a minimal separator). 
In \cite{ckl} it is shown that being both simplicial and admissible guarantees 
a vertex to be the end-vertex of an LBFS of a graph and moreover, 
the following two problems have been considered:

\begin{tabbing}
\= {\bf End-vertex problem:} \\

\> \hspace{6mm} \= {\bf Instance}:\ A graph $G$ on $n$ vertices with a specified vertex $t$.\\
\> \> {\bf Question}:\ Is there an LBFS ordering $\sigma$ of $G$ such $\sigma(n) = t$?
\end{tabbing}

\begin{tabbing}
\= {\bf Beginning-end-vertex problem:}\\

\> \hspace{6mm} \= {\bf Instance}:\ A graph $G$ on $n$ vertices with two specified vertices $s, t$.\\
\> \> {\bf Question}:\ Is there an LBFS ordering $\sigma$ of $G$ such that
               $\sigma(1) = s$ and $\sigma(n) = t$?
\end{tabbing}

The end-vertex problem is NP-complete for general graphs and remains NP-complete for 
the class of weakly chordal graphs, cf. \cite{ckl}. For the class of split graphs, 
the problem is solvable in polynomial time, cf.  \cite{chm}. 
The characterization of the end-vertices of LBFS of interval graphs
(as mentioned above) implies that the end-vertex problem is solvable in polynomial 
time for interval graphs. In fact, the end-vertex problem is shown to be
polynomial time solvable for the larger class of strongly chordal graphs,
cf. \cite{chm}. Despite many results on the (LBFS) end-vertices of chordal graphs
and of AT-free graphs are known, the end-vertex problem remains elusive for 
either class of graphs. There are few results on the beginning-end-vertex problem. 
Like the end-vertex problem, the beginning-end-vertex problem is NP-complete 
for weakly chordal graphs, cf. \cite{ckl}. 

In this paper, we consider the end-vertex problem as well as the beginning-end-vertex 
problem for bipartite graphs. It was left as an open problem in \cite{chm}
to determine the complexity of the end-vertex problem for bipartite graphs.
We prove that both the end-vertex and the begining-end-vertex problems for 
bipartite are NP-complete. We also study the end-vertex problem for AT-free 
bipartite graphs. 
We characterize the end-vertices of LBFS of AT-free bipartite graphs. Our 
characterization implies the end-vertex problem on AT-free bipartite graphs 
is solvable in polynomial time. It is known that every end-vertex $v$ of LBFS of 
an AT-free graph $G$ is admissible and has eccentricity differ by at most one
from the diameter of $G$. It is easy to show that each admissible vertex $v$
of an AT-free graph $G$ with eccentricity equal to the diameter of $G$ is 
the end-vertex of an LBFS. The complication arises when the eccentricity differs
from the diameter of the graph. We provide a simple condition for such a vertex 
to be the end-vertex of an LBFS. The class of AT-free bipartite graphs coincides 
with the class of proper interval graphs, cf. \cite{bls}. 
Our characterization of the end-vertices of LBFS of AT-free bipartite graphs may
be viewed as a result parallel to the characterization of the end-vertices of LBFS 
of interval graphs.

All graphs considered in this paper are simple (i.e., containing no loops or
multiple edges). Let $G$ be a graph and $v$ be a vertex $v$ in $G$. 
We use $N(v)$ to denote the {\em neighbourhood} and $N[v]$ ($=N(v) \cup \{v\}$) 
the {\em closed neighbourhood} of $v$. We say that a path $P$ {\em misses} 
vertex $v$ in $G$ if $P \cap N[v] \neq \emptyset$.
A {\em dominating path} in $G$ is a path $P$ such that 
no vertex in $G$ is missed by $P$. A pair of vertices $x, y$ is 
a {\em dominating pair} if every $(x,y)$-path is a dominating path in $G$.

An {\em asteroidal triple} in $G$ is an independent set of three vertices
such that between any two of the three vertices there is a path that misses
the third vertex. If $G$ does not contain an asteroidal triple then it is 
called {\em asteroidal triple-free} (AT-free). Two vertices $x, y$ of $G$ 
are called {\em unrelated with respect to} $z$ if there exists an 
$(x,z)$-path that misses $y$ and there is a $(y,z)$-path that misses $x$.
A vertex $z$ is {\em admissible} if there do not exist two vertices unrelated 
with respect $z$. According to \cite{cos99}, every LBFS-ordering $\sigma$ of 
an AT-free graph is an {\em admissible elimination ordering}, that is, 
for each $i = 1, 2, \dots, n$, $\sigma(i)$ is an admissible vertex in 
the subgraph induced by $\sigma(1), \sigma(2), \dots, \sigma(i)$; 
in particular, every AT-free graph has an admissible vertex.
Admissible vertices can be used to find dominating pairs and in fact,
every admissible vertex is a dominating pair vertex, cf. \cite{cos99}.

We use $d(x,y)$ to denote the {\em distance} between vertices $x, y$.  
The {\em diameter} of $G$, denoted by $\text{diam}(G)$, is the maximum distance 
of any two vertices. 
If $d(x,y) = \text{diam}(G)$, then we say that $x,y$ are a {\em diametrical
pair}. When $x, y$ are both diametrical and dominating, they are called
a {\em diametrical dominating pair}.

For a vertex $w$, we use $L_{\ell}(w)$ to denote the set of all
vertices $u$ with $d(u,w) = \ell$. The maximum value $\ell$ for which
$L_{\ell}(w) \neq \emptyset$ is called the {\em eccentricity} of $w$ and 
is denoted by $\text{ecc}(w)$. When $\ell = \text{ecc}(w)$, each vertex of 
$L_{\ell}(w)$ is called an {\em eccentric vertex} of $w$.

A graph is {\em chordal} if it does not contain an induced cycle of length
$\geq 4$. Every chordal graph has a {\em simplicial} vertex (i.e., $N(v)$ 
induces a clique). As mentioned above, every LBFS-ordering $\sigma$ of 
a chordal graph is a {\em perfect elimination ordering}, that is,
for each $i = 1, 2, \dots, n$, $\sigma(i)$ is a simplicial vertex in 
the subgraph of $G$ induced by $\sigma(1), \sigma(2), \dots, \sigma(i)$. 
A graph $G$ is an {\em interval graph} if there is a family of intervals 
$I_v, v \in V(G)$ such that two vertices $u, v$ are adjacent in $G$ if and only
if $I_u \cap I_v \neq \emptyset$. Interval graphs are exactly the AT-free 
chordal graphs, cf. \cite{lb}. 

We shall also call a bipartite graph a {\em bigraph}.
A bipartite graph $G$ with bipartition $(X,Y)$ is called an 
{\em interval bigraph} if there is a family of intervals $I_v, v \in X \cup Y$
in a line such that for all $x \in X$ and $y \in Y$, $x, y$ are adjacent 
in $G$ if and only if $I_x \cap I_y \neq \emptyset$, cf. \cite{hh,h}.
If the intervals can be chosen so that no interval properly contains 
another interval then $G$ is called a {\em proper interval bigraph}. 
Various characterizations of interval bigraphs and proper interval bigraphs
can be found in \cite{bls,hh,h}. In particular, 
proper interval bigraphs are precisely the AT-free bigraphs.

\section{General bigraphs}

The end-vertex problem and the begining-end-vertex problem are both NP-complete
for weakly chordal graphs, cf. \cite{ckl}. It was left as an open problem 
in \cite{chm} to determine the complexity of the end-vertex problem for bigraphs. 
In this section we will show that the end-vertex problem and 
the begining-end-vertex problem are both NP-complete for bigraphs. 

\begin{tm} \label{npc-bev}
The beginning-end-vertex problem is NP-complete for bigraphs.
\end{tm}

The proof of Theorem \ref{npc-bev} is similar to the one in \cite{chm} which shows 
that the (corresponding) beginning-end-vertex problem for BFS is NP-complete 
for bigraphs.

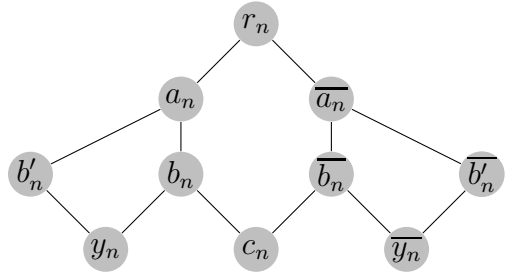
\begin{figure}[hb]
\centering
\begin{tikzpicture}
  \tikzstyle{vertex}=[circle,fill=black!25,minimum size=17pt,inner sep=0pt]
    \node[vertex] (a) at (0,0) {$r_n$};
    \node[vertex] (b) at (-1,-1) {$a_n$};
    \node[vertex] (c) at (1,-1) {$\overline{a_n}$};
    \node[vertex] (d) at (-1,-2) {$b_n$};
    \node[vertex] (e) at (1,-2) {$\overline{b_n}$};
    \node[vertex] (f) at (0,-3) {$c_n$};
    \node[vertex] (g) at (-3,-2) {$b_n'$};
    \node[vertex] (h) at (-2,-3) {$y_n$};
    \node[vertex] (i) at (3,-2) {$\overline{b_n'}$};
    \node[vertex] (j) at (2,-3) {$\overline{y_n}$};
   \draw (a) -- (b);
   \draw (a) -- (c);
   \draw (d) -- (b);
   \draw (f) -- (d);
   \draw (e) -- (c);
   \draw (e) -- (f);
   \draw (d) -- (h);
   \draw (g) -- (h);
   \draw (g) -- (b);
   \draw (i) -- (j);
   \draw (i) -- (c);
   \draw (j) -- (e);
\end{tikzpicture}
\caption{The graph $H_n$}
\label{hn}
\end{figure}

For every $n \in \mathbb{N}$, we define the graph $G_n$, with the special vertex 
$r_n$ called {\em the root}, recursively as follows:

\begin{itemize}
\item $G_0$ is the graph with one vertex $r_0$.
\item $G_n$ is obtained from $G_{n-1}$ and the graph $H_n$ in Figure \ref{hn} 
      by first adding an edge between the root $r_{n-1}$ of $G_{n-1}$ and the vertex
      $c_n$ in $H_n$ and then attaching to $y_n$ (respectively, $\overline{y_n}$) 
      an $(x_n,y_n)$-path (respectively, an $(\overline{x_n},\overline{y_n})$-path) 
      of length $4n-3$.
\end{itemize}

\begin{figure}[htbf]
\centering
\begin{tikzpicture}
  \tikzstyle{vertex}=[circle,fill=black!25,minimum size=17pt,inner sep=0pt]
    \node[vertex] (a) at (0,0) {$r_2$};
    \node[vertex] (b) at (-1,-1) {$a_2$};
    \node[vertex] (c) at (1,-1) {$\overline{a_2}$};
    \node[vertex] (d) at (-1,-2) {$b_2$};
    \node[vertex] (e) at (1,-2) {$\overline{b_2}$};
    \node[vertex] (f) at (0,-3) {$c_2$};
    \node[vertex] (g) at (-3,-2) {$b_2'$};
    \node[vertex] (h) at (-2,-3) {$y_2$};
    \node[vertex] (i) at (3,-2) {$\overline{b_2'}$};
    \node[vertex] (j) at (2,-3) {$\overline{y_2}$};
   \draw (a) -- (b);
   \draw (a) -- (c);
   \draw (d) -- (b);
   \draw (f) -- (d);
   \draw (e) -- (c);
   \draw (e) -- (f);
   \draw (d) -- (h);
   \draw (g) -- (h);
   \draw (g) -- (b);
   \draw (i) -- (j);
   \draw (i) -- (c);
   \draw (j) -- (e);

    \node[vertex] (aa) at (0,-4) {$r_1$};
    \node[vertex] (bb) at (-1,-5) {$a_1$};
    \node[vertex] (cc) at (1,-5) {$\overline{a_1}$};
    \node[vertex] (dd) at (-1,-6) {$b_1$};
    \node[vertex] (ee) at (1,-6) {$\overline{b_1}$};
    \node[vertex] (ff) at (0,-7) {$c_1$};
    \node[vertex] (gg) at (-3,-6) {$b_1'$};
    \node[vertex] (hh) at (-2,-7) {$y_1$};
    \node[vertex] (ii) at (3,-6) {$\overline{b_1'}$};
    \node[vertex] (jj) at (2,-7) {$\overline{y_1}$};
   \draw (aa) -- (bb);
   \draw (aa) -- (cc);
   \draw (dd) -- (bb);
   \draw (ff) -- (dd);
   \draw (ee) -- (cc);
   \draw (ee) -- (ff);
   \draw (dd) -- (hh);
   \draw (gg) -- (hh);
   \draw (gg) -- (bb);
   \draw (ii) -- (jj);
   \draw (ii) -- (cc);
   \draw (jj) -- (ee);

   \draw (aa) -- (f);
    \node[vertex] (aaa) at (0,-8) {$r_0$};
    \node[vertex] (bbb) at (2,-8) {$\overline{x_1}$};
    \node[vertex] (ccc) at (-2,-8) {$x_1$};
   \draw (aaa) -- (ff);
\draw (bbb) -- (jj);
   \draw (ccc) -- (hh);
    \node[vertex] (bbbb) at (2,-4) {$$};
    \node[vertex] (cccc) at (-2,-4) {$$};
    \node[vertex] (l1) at (4,-5) {$$};
    \node[vertex] (l2) at (4,-6) {$$};
    \node[vertex] (l3) at (4,-7) {$$};
    \node[vertex] (l4) at (4,-8) {$\overline{x_2}$};
   \draw (bbbb) -- (l1);
   \draw (l2) -- (l1);
   \draw (l2) -- (l3);
   \draw (l3) -- (l4);

    \node[vertex] (r1) at (-4,-5) {$$};
    \node[vertex] (r2) at (-4,-6) {$$};
    \node[vertex] (r3) at (-4,-7) {$$};
    \node[vertex] (r4) at (-4,-8) {$x_2$};
   \draw (cccc) -- (r1);
   \draw (r2) -- (r1);
   \draw (r2) -- (r3);
   \draw (r3) -- (r4);

   \draw (bbbb) -- (j);
   \draw (cccc) -- (h);

\end{tikzpicture}
\caption{The graph $G_2$.}
\label{g2}
\end{figure}
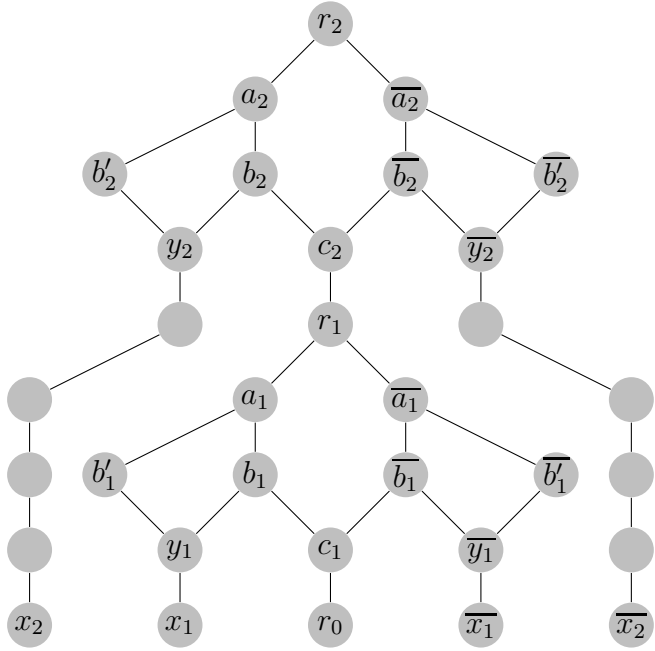

The graph $G_1$ is simply the one obtained from $H_1$ by adding three vertices
$x_1, r_0, \overline{x_1}$ adjacent to $y_1, c_1, \overline{y_1}$ respectively. 
The graph $G_2$ is depicted in Figure~\ref{g2}.

It is easy to verify that each $G_n$ is bipartite and has $4n^2 + 8n + 1$ vertices. 
Each vertex in $G_n$ is of distance at most $4n$ from the root $r_n$. 
The vertices of distance $4n$ from $r_n$ in $G_n$ are
$x_1, \dots, x_n, r_0, \overline{x_1}, \dots, \overline{x_n}$.
The following proposition will be useful in the proof of Theorem \ref{npc-bev}.

\begin{prp}\label{sort}
Let $\sigma$ be an LBFS ordering of $G_n$ with $\sigma(1) = r_n$. Then for each 
$1 \le i \le n$, either 
$\sigma^{-1}(x_i) < \sigma^{-1}(r_0) < \sigma^{-1}(\overline{x_i})$ or
$\sigma^{-1}(\overline{x_i}) < \sigma^{-1}(r_0) < \sigma^{-1}(x_i)$, that is
exactly one of $x_i$ and $\overline{x_i}$ is before $r_0$ in $\sigma$.
Moreover, each of the $2^n$ choices of one between $x_i$ and $\overline{x_i}$ 
for each $1 \leq i \leq n$ can be obtained as the set of vertices that appear 
before $r_0$ for some LBFS ordering of $G_n$.
\end{prp}
\pf We prove this by induction on $n$. When $n=0$, there is nothing to prove. 
Assume the statements hold for $G_{n-1}$ and consider $G_{n}$. 
Starting at $r_n$, the LBFS selects either $a_n$ or $\overline{a_n}$ to visit
next. Suppose that $a_n$ is visited before $\overline{a_n}$ 
(i.e., $\sigma^{-1}(a_n) < \sigma^{-1}(\overline{a_n})$). Then among
the four vertices of distance two from $r_n$, $b'_n$ and $b_n$ will be visited 
before either of $\overline{b_n}, \overline{b'_n}$. Hence the three vertices of 
distance three from $r_n$ must be visited in the order $y_n, c_n, \overline{y_n}$,
that is, $\sigma^{-1}(y_n) < \sigma^{-1}(c_n) < \sigma^{-1}(\overline{y_n})$.
Note that $y_n, c_n, \overline{y_n}$ belong to different components of 
$G_n - \{b'_n, b_n, \overline{b_n}, \overline{b'_n}\}$ that contain
$x_n, r_0, \overline{x_n}$ respectively. It follows that we must have
$\sigma^{-1}(x_n) < \sigma^{-1}(r_0) < \sigma^{-1}(\overline{x_n})$.
Similarly, if $\sigma^{-1}(\overline{a_n}) < \sigma^{-1}(a_n)$, then 
$\sigma^{-1}(\overline{x_n}) < \sigma^{-1}(r_0) < \sigma^{-1}(x_n)$.
Therefore exactly one of $x_n$ and $\overline{x_n}$ is before $r_0$ and any
one of the two vertices can appear before $r_0$ for some LBFS ordering of $G_n$.
The rest of the statements follow from the inductive hypothesis. 
\qed

{\bf Proof of Theorem \ref{npc-bev}:}
The proof uses a reduction from 3-$SAT$. 
Suppose that ${\cal I} = (x_1,\ldots,x_n; C_1,\ldots,C_m)$ is an instance of 3-$SAT$
where each $x_i$ with $1 \leq i \leq n$ is a variable and each $C_j$ with
$1 \leq j \leq m$ is a clause of size 3 over the variables and their negations.
We construct the graph $G_{\cal I}$ from $G_n$ (defined as above) by
adding $m+1$ new vertices $c_1, c_2, \dots, c_m, t$ in such a way that $t$ 
is adjacent only to $r_0$ and for each $1 \leq i \leq n$ and $1 \leq j \leq m$, 
$c_j$ is adjacent to $x_i$ (respectively, $\overline{x_i}$) if and only if 
$x_i$ (respectively, $\overline{x_i}$) is contained in the clause $C_j$.
We claim that $\cal I$ is satisfiable if and only if there is an LBFS ordering of 
$G_{\cal I}$ that begins at $r_n$ and ends at $t$. First note that every vertex 
in $G_{\cal I}$ is of distance at most $4n+1$ from $r_n$ and the vertices of 
distance $4n+1$ from $r_n$ are $c_1, c_2, \dots, c_m, t$.
Suppose that $\sigma$ is an LBFS ordering of $G_{\cal I}$ that begins at $r_n$ and 
ends at $t$. We assign a truth value to each variable as follows:
for each $x \in \{x_1, \dots, x_n\}$, $x$ is {\em true} if 
and only if $\sigma^{-1}(x) < \sigma^{-1}(r_0)$.
By Proposition \ref{sort}, exactly one of $x_i$ and $\overline{x_i}$
is assigned to be true for each $1 \leq i \leq n$.  
Since $\sigma^{-1}(c_j) < \sigma^{-1}(t)$ for each $1 \leq j \leq m$, each $c_j$ 
must be adjacent to a vertex that is before $r_0$ in $\sigma$, that is, 
at least one variable in $C_j$ is true. Hence $\cal I$ is satisfiable. 
Conversely, suppose that there is a truth value assignment to the variables that 
satisfies $\cal I$.
By Proposition \ref{sort}, there is an LBFS ordering of $G_{\cal I}$ that begins
at $r_n$ such that the vertex $x_i$ appears before $r_0$ if and only if
the corresponding variable $x_i$ is true for each $1 \leq i \leq n$. 
Since each $C_j$ contains at least one true variable, 
the vertex $c_j$ is adjacent to at least one vertex appear before $r_0$
in the LBFS ordering. Therefore the LBFS ordering must end at $t$. 
\qed

\begin{prp}\label{ev-bipart-reduc}
The beginning-end-vertex problem reduces in polynomial time to 
the end-vertex problem (for bigraphs).
\end{prp}
\pf Given a graph $G$ with two specified vertices $s, t$, we build $G'$ from $G$ 
by attaching to $s$ an $(s,s')$-path $P$ of length $\text{diam}(G)+1$. 
Clearly, $G'$ can be constructed in polynomial time 
(and when $G$ is bipartite so is $G'$). If some LBFS ordering of $G$ begins at 
$s$ and ends at $t$, then it is easy to see that there is an LBFS ordering of $G'$ 
that begins at $s'$ and ends at $t$.
Conversely, if some LBFS ordering of $G'$ that ends at $t$, then it must begin
at some vertex in $P$, which implies there is an LBFS ordering of $G$ that
begins at $s$ and ends at $t$.
\qed

Combining Theorems~\ref{npc-bev} and~\ref{ev-bipart-reduc} we have 
the following:

\begin{tm}\label{npc-ev}
The end-vertex problem is NP-complete for bigraphs.
\qed
\end{tm}

\section{AT-free bigraphs}

The goal of this section is to characterize the end-vertices of LBFS of 
AT-free bigraphs and to show that the end-vertex problem is solvable 
in polynomial time for AT-free bigraphs. To achieve this goal a few lemmas 
need to be in place first.

\begin{lemma} \cite{cdhp,cos99} \label{ad}
If $v$ is the end-vertex of an LBFS of an AT-free graph $G$, then $v$ is 
admissible and $\text{ecc}(v) \geq \text{diam}(G) - 1$.
\qed
\end{lemma} 

\begin{lemma} \cite{cos99} \label{dd}
Let $G$ be a connected AT-free graph $G$ and $v$ be an admissible vertex in $G$. 
Suppose that there is an LBFS ordering which begins at $v$ and ends at $w$.
Then $v, w$ are a dominating pair in $G$. 
Moreover, if $\text{ecc}(v) = \text{diam}(G)$, 
then $v, w$ are a diametrical dominating pair. 
\qed
\end{lemma}

Let $z$ be a vertex in graph $G$ and $\ell$ be a natural number. Recall that
$L_{\ell}(z)$ is the set of all vertices of distance $\ell$ from $z$. 
We shall use $N_{\ell}(a)$ to denote the set of all neighbours of $a$ 
in $L_{\ell}(z)$, that is, $N_{\ell}(a) = N(a) \cap L_{\ell}(z)$.

\begin{lemma} \label{comparable}
Let $G$ be an AT-free bigraph and $z$ be a vertex of $G$.
Suppose that $C$ is a connected component of $G - N[z]$ and that
$a, b \in L_{\ell}(z)$ are two vertices in $C$. Then
\begin{enumerate}
\item $N_{\ell-1}(a) \subseteq N_{\ell-1}(b)$ or
      $N_{\ell-1}(a) \supseteq N_{\ell-1}(b)$;
\item $N_{\ell+1}(a) \subseteq N_{\ell+1}(b)$ or
      $N_{\ell+1}(a) \supseteq N_{\ell+1}(b)$;
\item $N_{\ell-1}(a) \subseteq N_{\ell-1}(b)$ if and only if
      $N_{\ell+1}(a) \supseteq N_{\ell+1}(b)$.
\end{enumerate}
\end{lemma}
\pf Since $a, b \notin N[z]$, $\ell \geq 2$.  
Suppose that $N_{\ell-1}(a) \not\subseteq N_{\ell-1}(b)$ 
and $N_{\ell-1}(a) \not\supseteq N_{\ell-1}(b)$.  Then there exist
$a' \in N_{\ell-1}(a) \setminus N_{\ell-1}(b)$ and
$b' \in N_{\ell-1}(b) \setminus N_{\ell-1}(a)$. We claim that $z, a, b$ form
an asteroidal triple, a contradiction to the assumption that $G$ is AT-free. 
Indeed, any $(a,b)$-path in $C$ misses $z$, any shortest $(a,z)$-path containing 
$a'$ misses $b$ and similarly any shortest $(b,z)$-path containing $b'$ misses $a$.
This proves statement~1. Statement~2 follows from statement~1.

For statement~3, suppose to the contrary that
$N_{\ell-1}(a) \subsetneq N_{\ell-1}(b)$ and 
$N_{\ell+1}(a) \subsetneq N_{\ell+1}(b)$.
Then there exist $b' \in N_{\ell-1}(b) \setminus N_{\ell-1}(a)$ and
$b'' \in N_{\ell+1}(b) \setminus N_{\ell+1}(a)$. 
By statemen~1, there exists $a' \in N_{\ell-1}(a) \cap N_{\ell-1}(b)$.
We show that $G$ contains an asteroidal triple, which is a contradiction.
Suppose that $\ell = 2$. Then $z$ is adjacent to both $a', b'$. 
By statement~2 there exists $a'' \in N_{\ell+1}(a) \cap N_{\ell+1}(b)$. 
We obtain an asteroidal triple $\{z, a'', b''\}$: $za'aa''$ is a path missing $b''$,
$ab'bb''$ is a path missing $a''$, and $a''bb''$ is a path missing $z$. 
Suppose that $\ell > 2$. Then again by statement~1 (applied to $a', b'$) there 
exists $z' \in N_{\ell-2}(a') \cap N_{\ell-2}(b')$. Let $z''$ be any vertex
in $N_{\ell-3}(z')$. We obtain an asteroidal triple $\{a, b'', z''\}$:
$aa'bb''$ is a path missing $z''$, $aa'a'z''$ is a path missing $b''$, and
$b''bb'z'z''$ is a path missing $a$.    
\qed

Suppose that $C$ is a component of $G \setminus N[z]$ and $a, b \in N(z)$.
It follows from Lemma \ref{comparable} that either 
$N(a) \cap C \subseteq N(b) \cap C$ or $N(a) \cap C \supseteq N(b) \cap C$.
In particular, if $c \in N(z)$ is a vertex adjacent to the maximum number of 
vertices in $C$, then for any $u \in L_{\ell}(z) \cap C$ with $\ell \geq 2$,
$d(c,u) \leq \ell - 1$.
We call $C$ a {\em deep} component of $G \setminus N[z]$ if it contains 
an eccentric vertex of $z$. Note that a deep component of $G-N[z]$ 
exists when and only when $\text{ecc}(z) \geq 2$.

\begin{lemma} \label{2deeps}
Let $G$ be a connected AT-free bigraph and $z$ be a vertex of $G$.
If $\text{ecc}(z) \geq 3$, then $G - N[z]$ has at most two deep components.
\end{lemma}
\pf Suppose that $G - N[z]$ has three or more deep components. Let
$a, b, c$ be eccentric vertices of $z$ belonging to three different deep
components of $G - N[z]$. Then $d(a,z) = d(b,z) = d(c,z) = \text{ecc}(z) \geq 3$.
Since $a, b, c$ belong to three different deep components of $G - N[z]$,
each of $a, b, c$ is joined to $z$ by a path that misses the other two vertices.
It follows that there is a path joining any pair of $a, b, c$ that misses
the third vertex, i.e., $\{a, b, c\}$ is an asteroidal triple, a contradiction
to the assumption that $G$ is AT-free.
\qed

We remark that, when $eec(z) = 2$, $G - N[z]$ can have any number of deep components
as each vertex in $L_2(z)$ forms one of those.

\begin{lemma} \label{ddp}
Let $G$ be a connected AT-free bigraph and $v$ be an admissible vertex with 
$\text{ecc}(v) = \text{diam}(G) - 1$. 
Suppose that $x, y$ are a diametrical dominating pair.
Then $v$ is adjacent to one of $x, y$.
Moreover, there is a shortest $(x,y)$-path containing $v$.
\end{lemma}
\pf Let $k = \text{diam}(G)$. Suppose that $v$ is adjacent to neither of $x, y$.  
Let $P:\ xx_1x_2\dots x_{k-1}y$ be a shortest $(x,y)$-path. 
We claim that $v$ is not in $P$. Indeed, $v \neq x_1$ and $v \neq x_{k-1}$ 
as $v$ is adjacent to neither of $x, y$. If $v = x_i$ for some $1 < i < k-1$, then
$x, y$ are unrelated vertices with respect to $v$;
$xx_1 \dots x_i$ is an $(x,v)$-path that misses $y$ and $yx_{k-1} \dots x_i$
is a $(y,v)$-path that misses $x$, a contradiction to the assumption that 
$v$ is admissible. Hence $v$ is not in $P$. The arbitrary choice of $P$ implies
that $v$  is not in any shortest $(x,y)$-path. Since $x, y$ are a dominating pair, 
$v$ is adjacent to $x_j$ for some $1 \leq j \leq k-1$. 
If $j \neq 1$ and $j \neq k-1$, then $xx_1 \dots x_jv$ is an $(x,v)$-path 
that misses $y$ and $yx_{k-1}\dots x_jv$ is a $(y,v)$-path that misses $x$,
a contradiction. So $j = 1$ or $j = k-1$. Assume without loss of generality 
that $j = k-1$. Then $d(v,y) = 2$. Since $\text{ecc}(v) = k-1$, $d(x,v) \leq k-1$.
If $d(x,v) \leq k-2$, then $k = d(x,y) \leq d(x,v) + d(v,y) \leq k-2+2 = k$, which
implies that $v$ is contained a shortest $(x,y)$-path, a contradiction.
So $d(x,v) = k-1$. Now any $(x,v)$-path of length $k-1$ along with 
the path $xx_1 \dots x_{k-1}v$ form a closed walk of length $2k-1$. This 
contradicts the assumption that $G$ is bipartite.
Therefore $v$ is adjacent to one of $x, y$.

To prove that there is a shortest $(x,y)$-path containing $v$, we assume 
by symmetry that $v$ is adjacent to $y$.  
Then $d(x,y) \leq d(x,v) + 1 \leq \text{ecc}(v) + 1 = k-1 + 1 = d(x,y)$. This 
implies that there is a shortest $(x,y)$-path containing $v$. 
\qed

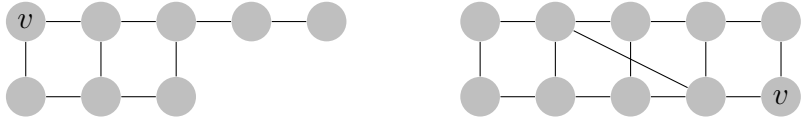
\begin{figure}[ht]
\centering
\begin{tikzpicture}
  \tikzstyle{vertex}=[circle,fill=black!25,minimum size=15pt,inner sep=0pt]
    \node[vertex] (a) at (0,0) {$$};
    \node[vertex] (b) at (-1,0) {$$};
    %\node[vertex] (d) at (1,0) {$w$};
    \node[vertex] (d) at (1,0) {$$};
    \node[vertex] (e) at (0,1) {$$};
    \node[vertex] (f) at (-1,1) {$v$};
    \node[vertex] (h) at (1,1) {$$};
    \node[vertex] (i) at (2,1) {$$};
    \node[vertex] (j) at (3,1) {$$};
   \draw (a) -- (b);
   \draw (d) -- (a);
   \draw (a) -- (e);

\draw(i)--(j);
\draw(h)--(i);

   \draw (h) -- (e);
   \draw (d) -- (h);
   \draw (f) -- (b);

   \draw (f) -- (e);

\end{tikzpicture}
$~~~~~~~~~$
\begin{tikzpicture}
  \tikzstyle{vertex}=[circle,fill=black!25,minimum size=15pt,inner sep=0pt]
    \node[vertex] (a) at (0,0) {$$};
    \node[vertex] (b) at (-1,0) {$$};
    \node[vertex] (d) at (1,0) {$$};
    \node[vertex] (e) at (0,1) {$$};
    \node[vertex] (f) at (-1,1) {$$};
    \node[vertex] (h) at (1,1) {$$};
    %\node[vertex] (h) at (1,1) {$w$};
    \node[vertex] (i) at (2,1) {$$};
    \node[vertex] (j) at (3,1) {$$};
    \node[vertex] (k) at (3,0) {$v$};
    \node[vertex] (l) at (2,0) {$$};

   \draw (a) -- (b);
   \draw (d) -- (a);
   \draw (a) -- (e);

\draw(i)--(j);
\draw(h)--(i);

\draw(k)--(j);
\draw(l)--(k);

\draw(d)--(l);
\draw(i)--(l);
\draw(e)--(l);
   \draw (h) -- (e);
   \draw (d) -- (h);
   \draw (f) -- (b);

   \draw (f) -- (e);

\end{tikzpicture}
\caption{Two AT-free bigraphs in which $v$ satisfies 
$\text{ecc}(v) = \text{diam}(G) - 1$ but is not the end-vertex of any LBFS.}
\label{annoying}
\end{figure}

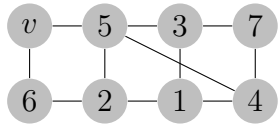
\begin{figure}[ht]
\centering
\begin{tikzpicture}
  \tikzstyle{vertex}=[circle,fill=black!25,minimum size=17pt,inner sep=0pt]
    \node[vertex] (a) at (0,0) {$1$};
    \node[vertex] (b) at (-1,0) {$2$};
    \node[vertex] (c) at (-2,0) {$6$};
    \node[vertex] (d) at (1,0) {$4$};
    \node[vertex] (e) at (0,1) {$3$};
    \node[vertex] (f) at (-1,1) {$5$};
    \node[vertex] (g) at (-2,1) {$v$};
    \node[vertex] (h) at (1,1) {$7$};

   \draw (a) -- (b);
   \draw (a) -- (d);
   \draw (a) -- (e);
   \draw (d) -- (f);
   \draw (h) -- (e);
   \draw (d) -- (h);
   \draw (f) -- (b);

   \draw (f) -- (e);
   \draw (g) -- (f);
   \draw (g) -- (c);
   \draw (c) -- (b);

\end{tikzpicture}
\caption{An AT-free bigraph in which $v$ satisfies 
$\text{ecc}(v) = \text{diam}(G) - 1$ and is the end-vertex of an LBFS.}
\label{fig:ecc-end-ce}
\end{figure}

Figure~\ref{annoying} shows two AT-free bigraphs in which the vertex $v$ satisfies
$\text{ecc}(v) = \text{diam}(G) - 1$ but is not the end-vertex of any LBFS. 
Figure~\ref{fig:ecc-end-ce} shows an AT-free bigraph in which $v$  satisfies 
$\text{ecc}(v) = \text{diam}(G) - 1$ and is the end-vertex of an LBFS; 
the numbering is an LBFS ordering.  

\begin{tm} \label{main}
Let $G$ be a connected AT-free bigraph and $v$ be a vertex of $G$. Then
$v$ is the end-vertex of an LBFS if and only if there exists a vertex $w$ such that, 
for every eccentric vertex $u$ of $w$, $N(v) \subseteq N(u)$.
\end{tm}
\pf Suppose that there exists a vertex $w$ such that, for every eccentric vertex $u$ 
of $w$, $N(v) \subseteq N(u)$. If $G$ has only one vertex, then $v$ is clearly 
the end-vertex of an LBFS. So assume that $G$ has at least two vertices. 
If $v = w$, then $\text{ecc}(w) = 2$ and $G$ is a complete bigraph. In this case, 
let $w'$ be any eccentric vertex of $w$, we have $N(v) = N(u)$ for every eccentric 
vertex $u$ of $w'$. Thus $w'$ can be used in the placec of $w$. 
So we may assume that $v \neq w$. 
Apply LBFS begining at $w$. It must end at an eccentric vertex of $w$. 
Since $N(v) \subseteq N(u)$ for every eccentric vertex $u$ of $w$, $v$ is 
an eccentric vertex of $w$ and it is possible to have it being visited last. 

For the other direction suppose that $v$ is an LBFS end-vertex. 
Let $k = \text{diam}(G)$.
By Lemma \ref{ad}, $v$ is admissible and $\text{ecc}(v) \geq k-1$. 
We show there exists a vertex $w$ such that $N(v) \subseteq N(u)$ for every 
eccentric vertex $u$ of $w$. When $k \leq 2$, $G$ is a complete bigraph. 
Giving the fact that $v$ is an LBFS end-vertex it is easy to see such
a vertex $w$ exists. So we assume $k \geq 3$.  
We consider two cases:

{\bf Case~1.} $\text{ecc}(v) = k$.

Let $w$ be any vertex with $d(v,w) = k$. Clearly $v$ is an eccentric 
vertex of $w$. Suppose there exists an eccentric vertex $u$ of $w$ such
that $N(v) \not\subseteq N(u)$. If $u$ and $v$ are in the same component of 
$G - N[w]$ then by Lemma \ref{comparable}, $N(u) \subsetneq N(v)$ and $v$ 
cannot be an LBFS end-vertex, contradicting the assumption.
Hence $u$ and $v$ belong to different components of $G - N[w]$. 
But then we have $d(u,v) \geq 2(k-1) > k$, a contradiction.
 
{\bf Case~2.} $\text{ecc}(v) = k-1$.

Let $\sigma$ be an LBFS ordering with $\sigma(n) = v$ and let $z = \sigma(1)$. 
Clearly, $v$ is an eccentric vertex of $z$. If $N(v) \subseteq N(u)$ for every
eccentric vertex $u$ of $z$ then we are done. So assume that this is not the case;
there is an eccentric vertex $u$ of $z$ such that $N(v) \not\subseteq N(u)$. 
The existence of such a vertex $u$ implies $\text{ecc}(z) \geq 2$.

Consider first the case when $\text{ecc}(z) = 2$. Then $k \leq 4$ as any two 
vertices are joined by a path of length $\leq 4$ (through the vertex $z$).
Let $A = N(v) \setminus N(u)$, $B = N(u) \setminus N(v)$, and $C = N(u) \cap N(v)$.
Since $N(v) \not\subseteq N(u)$, $A \neq \emptyset$. Since $v$ is an LBFS 
end-vertex, $B \neq \emptyset$. If $C = \emptyset$, then 
$d(u,v) \geq 4$ and hence 
\[\text{ecc}(v) \geq d(u,v) \geq 4 \geq k = \text{ecc}(v)+1,\]
a contradiction. Thus $C \neq \emptyset$. 
Clearly, $(A \cup B \cup C) \subseteq N(z)$. In fact we must have
$A \cup B \cup C  = N(z)$ as otherwise any vertex in 
$N(z) \setminus (A \cup B \cup C)$ would form an asteroidal triple 
with $u, v$, a contradiction to the assumption that $G$ is AT-free.
It is easy to see that the distance between $v$ and any vertex in $B$ is 3. 
This implies that $k = 4$ and hence $\text{ecc}(v) = k-1 = 3$. 
Let $x, y$ be any pair of diametrical vertices.  
Clearly, $x, y$ are both eccentric vertices of $z$. 
Each of $x, y$ must have a neighbour in $A \cup B$ as otherwise its distance to 
$v$ is 4, contradicting the fact that $\text{ecc}(v) = 3$. 
Let $x', y' \in A \cup B$ be the neighbours of $x, y$ respectively. Then
$xx'v$ is an $(x,v)$-path missing $y$ and $yy'v$ is a path missing $x$. 
Hence $x, y$ are unrelated vertices with respect to $v$, which contradicts
the fact that $v$ is admissible.
So from now on we may assume that $\text{ecc}(z) \geq 3$. 

By Lemma \ref{comparable}, $u$ and $v$ belong to different components of 
$G - N[z]$. Denote by $C_1, C_2$ the two components of $G - N[z]$ which contain
$u, v$ respectively. Since $u, v$ are both eccentric vertices of $z$ belonging to
the different components $C_1, C_2$ of $G - N[z]$, $d(u,v) \geq 2(\text{ecc}(z)-1)$ 
and $C_1, C_2$ are deep components of $G - N[z]$. In view of Lemma \ref{2deeps}, 
$C_1, C_2$ are the only deep components of $G - N[z]$. 
Let $x, y$ be a diametrical dominating pair in $G$ which exists according to
Lemma \ref{dd}. Then by Lemma \ref{ddp}, $v$ is adjacent to one of $x, y$. 
Assume by symmetry that $v$ is adjacent to $y$, which implies 
that $y$ is also in $C_2$ and $d(z,y) = d(z,v) - 1 = \text{ecc}(z) - 1$. 
We show (by contradiction) that $x$ is in $C_1$ and is an eccentric vertex 
of $z$. Indeed, if $x$ is not an eccentric vertex of $z$, then 
$d(x,z) \leq \text{ecc}(z) -1$; if $x$ is in $C_2$, let $c \in N(z)$ be a vertex 
adjacent to the maximum number of vertices in $C_2$, then 
$d(x,c) \leq \text{ecc}(z) - 1$ and $d(c,y) \leq \text{ecc}(z) - 2$ (see the remarks
following Lemma \ref{comparable}).
In the former case, we have $d(x,y) \leq d(x,z) + d(z,y) \leq 2\text{ecc}(z)-2$ and
in the latter case, we have $d(x,y) \leq d(x,c) + d(c,y) \leq 2\text{ecc}(z)-3$.
Hence
\[\text{ecc}(v)+ 1 = k = d(x,y) \leq 2(\text{ecc}(z)-1) \leq d(u,v) 
           \leq \text{ecc}(v),\]   
which is a contradiction. 
Therefore $x$ is in $C_1$ and is an eccentric vertex of $z$. 
Since $x, y$ is a dominating pair, for every eccentric vertex $b$ of $z$ in $C_1$ 
we must have $N(b) \supseteq N(x)$.

Consider a shortest $(x,y)$-path that contains $v$, which exists according to
Lemma \ref{ddp}. Let $P:\ xx_1x_2\dots x_{k-2}vy$ be such a path. 
Then $P$ must contain a vertex in $N(z)$ as $x$ and $y$ belong to the different 
components $C_1, C_2$ of $G - N[z]$ respectively.
Let $x_{\alpha} \in N(z)$ be the vertex in $P$ with the smallest 
subscript and let $Q$ be the subpath $x_{\alpha}x_{\alpha+1} \dots vy$ of $P$. 
Since $\text{ecc}(z) - 1 \leq d(x_{\alpha},v) = d(x_{\alpha},y) - 1 
   \leq \text{ecc}(z)$ and 
$d(x_{\alpha},v)$ cannot be $\text{ecc}(z)$, we must have
$d(x_{\alpha},v) = \text{ecc}(z) - 1 = d(x_{\alpha},y) - 1$. 
Hence the length $Q$ is exactly $\text{ecc}(z)$. 
It follows that $P$ does not contain $z$ and moreover, if $Q$ is replaced by 
an $(x_{\alpha},y)$-path of length $\text{ecc}(z)$ through $z$ then we obtain 
another shortest $(x,y)$-path $P'$ containing $z$ but not $v$. 
The existence of the shortest $(x,y)$-path $P'$ (containing $z$) further implies 
the length of $xx_1x_2 \dots x_{\alpha}$ is $\text{ecc}(z) - 1$. 
Therefore we know that $k = 2\text{ecc}(z) - 1 = 2\alpha+1$.

Let $y'$ be any vertex in $N(z)$ with $d(y',y) = \text{ecc}(z) - 2$. Then 
$x_{\alpha-1}$ and $y'$ are not adjacent as otherwise replacing the subpath $Q$ of 
$P$ by any $(y',y)$-path of length $\text{ecc}(z) - 2$ we obtaining 
an $(x,y)$-path shorter than $P$, a contradiction to the fact that $P$ is 
a shortest $(x,y)$-path. 
For the same reason $x_{\alpha}$ is not adjacent to any vertex in 
a shortest $(y',y)$-path. Since $N(x_{\alpha}) \cap C_2$ and $N(y') \cap C_2$ are 
comparable (see the remarks following Lemma \ref{comparable}), 
$y'$ must be adjacent to $x_{\alpha+1}$.
Thus we have two vertices $x_{\alpha}, y' \in N(z)$, both adjacent to $x_{\alpha+1}$
and only $x_{\alpha}$ adjacent to $x_{\alpha-1}$.   

Let $A$ be the set of all vertices $a \in L_2(z) \cap C_1$ with
$\sigma^{-1}(a) < \sigma^{-1}(x_{\alpha+1})$ and $d(a,x) = \text{ecc}(z) - 2$.
Since $\sigma^{-1}(x) < \sigma^{-1}(v)$, $A \neq \emptyset$. 
No vertex $a$ in $A$ is adjacent to $y'$ as otherwise any $(x,a)$-path of 
length $\text{ecc}(z)-2$ and any $(y',y)$-path of length $ecc(z)-2$ together with 
$ay'$ form an $(x,y)$-path of length $2\text{ecc}(z)-3$, a contradiction to 
$d(x,y) = k = 2\text{ecc}(z) - 1$. 
Furthermore, every vertex $a \in A$ has a neighbour in $N(z)$ which is not 
a neighbour of $x_{\alpha+1}$ since $\sigma^{-1}(a) < \sigma^{-1}(x_{\alpha+1})$.

Let $w$ be a vertex in $N(z)$ that is a neighbour of some vertex $a \in A$ but 
not a neighbour of $x_{\alpha+1}$. 
We show that the vertex $w$ satisfies the properties desired by the theorem.
First it is easy to see that $d(w,v) = \text{ecc}(z) + 1$ ($= \frac{1}{2}(k+3)$ and
$wzx_{\alpha}x_{\alpha+1} \dots x_{k-2}v$ is a shortest $(w,v)$-path).
Consider an arbitrary vertex $b$ of $G$.
If $b$ is not an eccentric vertex of $z$, then $d(z,b) \leq \text{ecc}(z) - 1$ and
hence $d(w,b) \leq d(w,z) + d(z,b) = 1 + \text{ecc}(z) - 1 = \text{ecc}(z)$.
If $b \in C_1$ is an eccentric vertex of $z$, then from the above we know 
that $N(b) \supseteq N(x)$ and so 
$d(w,b) = d(w,x) = \text{ecc}(z) - 1$.
If $b \in C_2$ is an eccentric vertex of $z$, then
$d(w,b) \leq d(w,z) + d(z,b) = eec(z) + 1$.
Thus each eccentric vertex $b$ of $w$ is an eccentric vertex of $z$ in $C_2$
and hence we must have $N(v) \subseteq N(b)$.
This completes the proof.
\qed

Since the necessary and sufficent condition in Theorem \ref{main} 
for a vertex to be the end-vertex of an LBFS can be verified in polynomial time, 
we have the following:

\begin{tm}
The end-vertex problem for AT-free bigraphs is polynomial time solvable.
\qed
\end{tm}

\section{Concluding remarks}

We have proved in this paper that the end-vertex problem and the begining-end-vertex
problem are both NP-complete for bigraphs and that the end-vertex problem is 
polynomial time solvable for AT-free bigraphs. The NP-completeness result solves 
an open problem from \cite{chm}. The result on AT-free bigraphs follows from
a characterization of the end-vertices of LBFS of AT-free bigraphs obtained 
also in this paper. With a slight modification of the reductions in the proofs
Theorems \ref{npc-bev} and \ref{npc-ev} one can show that  
the beginning-end-vertex problem is NP-complete for bigraphs of maximum degree
three and the end-vertex problem is NP-complete for bigraphs of maximum degree
four.

The end-vertex problem and the beginning-end-vertex problem for other graph
search algorithms have been studied in \cite{chm,gorzny-thesis}. We state some 
of the results obtained in \cite{gorzny-thesis}: The end-vertex problem 
for Depth First Search (DFS) is NP-complete for bigraphs.
The end-vertex problem for Lexicographic Depth First Search (LDFS) is 
also NP-complete for chordal graphs.
Each of these solves an open problem in \cite{chm}.
The begining-end-vertex problem for BFS is polynomial time solvable 
for split graphs. This follows from a characterization of pairs $s, t$
in split graphs for which some BFS begins at $s$ and ends at $t$, which is 
a slight refinement of a result obtained in \cite{chm}.

\end{document}